\documentclass[twocolumn,superscriptaddress,prd,showkeys,showpacs,nofootinbib]{revtex4-1}
\usepackage{graphics}
\usepackage[colorlinks = true,
            linkcolor = cyan,
            urlcolor  = blue,
            citecolor = red,
            anchorcolor = blue]{hyperref}\usepackage{color}
\usepackage{amssymb}
\usepackage[nointegrals]{wasysym}
\usepackage{amsthm}
\usepackage{textcomp}
\usepackage{mathtools}
\usepackage{lineno}

\usepackage{comment}
\usepackage[separate-uncertainty,retain-explicit-plus,per-mode = symbol]{siunitx}
\usepackage{url}
\newcolumntype{P}[1]{>{\centering\arraybackslash}p{#1}}
\newcolumntype{C}[1]{>{\centering\arraybackslash}m{#1}}

\begin{document}
%\setpagewiselinenumbers
%\linenumbers
%-----------------------------------------------------------------------
%-----------------------------------------------------------------------

\title{Thermodynamic Stability of Xenon-Doped Liquid Argon Detectors}

%\noaffiliation%\collaboration{A Collaboration}%
%\documentclass[superscriptaddress]{revtex4-1}
%\begin{document}

\author{E. P. Bernard}  \email[Corresponding author, ] {bernard5@llnl.gov} 
\affiliation{Lawrence Livermore National Laboratory, 7000 East Ave., Livermore, CA 94550, USA}
\author{E. Mizrachi}
\affiliation{Lawrence Livermore National Laboratory, 7000 East Ave., Livermore, CA 94550, USA}
\affiliation{University of Maryland, Department of Physics, Campus Dr., College Park, MD 20742, USA}
\author{J. Kingston}
\affiliation{Lawrence Livermore National Laboratory, 7000 East Ave., Livermore, CA 94550, USA}
\affiliation{University of California Davis, Department of Physics, One Shields Ave., Davis, CA 95616, USA}
\author{J. Xu }
\affiliation{Lawrence Livermore National Laboratory, 7000 East Ave., Livermore, CA 94550, USA}
\author{S. V. Pereverzev}
\affiliation{Lawrence Livermore National Laboratory, 7000 East Ave., Livermore, CA 94550, USA}
\author{T. Pershing}
\affiliation{Lawrence Livermore National Laboratory, 7000 East Ave., Livermore, CA 94550, USA}
\author{R. Smith}
\affiliation{University of California Berkeley, Department of Physics, 366 Physics North, Berkeley, CA 94720, USA}
\author{C. G. Prior}
\affiliation{Duke University, Department of Physics, 120 Science Dr., Durham, NC 27708, USA}
\author{N. S. Bowden}
\affiliation{Lawrence Livermore National Laboratory, 7000 East Ave., Livermore, CA 94550, USA}
\author{A. Bernstein}
\affiliation{Lawrence Livermore National Laboratory, 7000 East Ave., Livermore, CA 94550, USA}
\author{C. R. Hall}
\affiliation{University of Maryland, Department of Physics, Campus Dr., College Park, MD 20742, USA}
\author{E. Pantic}
\affiliation{University of California Davis, Department of Physics, One Shields Ave., Davis, CA 95616, USA}
\author{M. Tripathi}
\affiliation{University of California Davis, Department of Physics, One Shields Ave., Davis, CA 95616, USA}
\author{D. N. McKinsey}
\affiliation{University of California Berkeley, Department of Physics, 366 Physics North, Berkeley, CA 94720, USA}
\author{P. S. Barbeau}
\affiliation{Duke University, Department of Physics, 120 Science Dr., Durham, NC 27708, USA}
%\maketitle
%\end{document}

%-----------------------------------------------------------------------
%-----------------------------------------------------------------------

\date{\today}% It is always \today, today,

%-----------------------------------------------------------------------
%-----------------------------------------------------------------------

\begin{abstract}
Liquid argon detectors are employed in a wide variety of nuclear and particle physics experiments.  The addition of small quantities of xenon to argon modifies its scintillation, ionization, and electroluminescence properties and can improve its performance as a detection medium. However, a liquid argon-xenon mixture can develop instabilities, especially in systems that require phase transitions or that utilize high xenon concentrations. In this work, we discuss the causes for such instabilities and describe a small (liter-scale) apparatus with a unique cryogenic circuit specifically designed to handle argon-xenon mixtures.  The system is capable of condensing argon gas mixed with $\mathcal{O}(1\%)$ xenon by volume and maintains a 
stable liquid mixture near the xenon saturation limit while actively circulating it in the gas phase. 
We also demonstrate control over instabilities that develop when the detector condition is allowed to deviate from optimized settings. 
This progress enables future liquid argon detectors to benefit from the effects of high concentrations of xenon doping, such as more efficient detection of low-energy ionization signals. This work also develops tools to study and mitigate instabilities in large argon detectors that use low concentration xenon doping. 
\end{abstract}

%-----------------------------------------------------------------------
%-----------------------------------------------------------------------
\maketitle
%\tableofcontents

%-----------------------------------------------------------------------
% Core of the document
%-----------------------------------------------------------------------

\section{Introduction}
\label{sec:intro}

Liquid argon plays a 
critical role in particle detection in a wide variety of nuclear and particle physics experiments, including detectors for hadron colliders \cite{Hervas2005}, neutrinos \cite{majumdar2021}, neutrinoless double beta decay \cite{gerdaphasetwo} and dark matter \cite{Darkside50,Darkside20kPDP}.  Liquid argon 
detectors can be scaled to large size \cite{PDuneSP_overview,PDune_DP_overview} and can be made both chemically and radioactively pure~\cite{Darkside20kCleanAr,Jingkephd}. Detectors using liquid argon as the target medium can sense both 
scintillation light and ionization electrons. The detection of liquid argon scintillation is usually achieved
indirectly by converting 128~nm UV light to visible wavelengths by use of fast fluorescent coatings such as 1,1,4,4-tetraphenyl-1,3-butadiene (TPB)~\cite{WLSSurvey}. This light is then sensed by photomultiplier tubes (PMTs) or silicon photomultipliers (SiPMs). 
Charge signals can be collected by drifting the ionization electrons in single phase argon time projection chambers (TPCs) through crossed wire planes or, in some newer schemes, to arrays of small individually instrumented anodes \cite{duneneardetCDR}.  Dual-phase TPCs can be made far more sensitive to low-energy ionization signals by drifting electrons to the liquid argon surface, extracting them, and amplifying the signal when the electrons drift through a short ($\sim$~1~cm) gas gap toward an anode wire or plane.  With gain fields of a few kV/cm, $\mathcal{O}(100)$ photons can be produced for each electron transiting the gap and individual extracted electrons can be resolved \cite{Darkside50}. This low-energy sensitivity, in combination with the recent observation of coherent neutrino nucleus scattering (CE$\nu$NS) in liquid argon \cite{akimov2021}, opens a new set of applications for liquid argon detectors optimized to sense the smallest signals. These include measuring non-standard neutrino interactions \cite{Miranda2020,Flores2020,Blanco2020}, core-collapse supernova neutrinos \cite{Darkside20kCCSN}, CNO solar neutrinos \cite{Abdullah2022} and non-intrusive monitoring of nuclear reactor fuel cycles \cite{Hagmann2004,Sangiorgio2012}.

The effects of doping argon with small ($<$1000~ppm\footnote{Concentrations in this paper are given in mole/mole}) concentrations of xenon have been explored by several teams over the last decade.  
Although argon with very low (typically incidental) xenon concentrations can absorb the 128~nm Ar$_2$ emission wavelength without re-emission~\cite{Neumeier2015B,Neumeier2015C}, carefully controlled xenon doping in liquid argon has been demonstrated to improve the detection of primary scintillation light.
The most pronounced effect is the transfer of energy from argon excitations to ArXe and (predominantly) Xe$_2$ dimers, which release light at 149~nm and 174~nm wavelengths upon dissociation.  The energy transfer is efficient for Xe concentrations above 10~ppm \cite{Cheshnovsky1972,Neumeier2015A}, and 
the 174~nm wavelength becomes predominant.    
The immediate benefit of the energy transfer to 174~nm is its compatibility with quartz-windowed sensing optics, especially PMTs, and with common UV reflectors such as polytetrafluoroethylene (PTFE).  
Large detectors also benefit from the shift of scintillation light to longer wavelengths due to the correspondingly longer Rayleigh scattering length in liquid argon~\cite{Grace2017,sotooton2021}, although it was reported that very high xenon concentrations shorten the scattering length~\cite{Seidel2002,Ishida1997}. 
Furthermore, the addition of xenon reduces liquid argon's sensitivity to scintillation quenching by impurities such as N$_2$ because xenon competes with these impurities in reacting with the long-lived Ar$_2$ triplet and releases the acquired energy through photon emission \cite{Kubota1982Nitrogenquench,Jones2013,Acciarri2010}.  
It was reported that xenon doping above 100~ppm increases the light yield of liquid argon at zero electric field by about a factor of 20--30\%, but the precise relationship between doping level and total photon yield is still being actively studied \cite{Fields2022,Segreto2021,Suzuki1993LightOA}.  

Xenon doping also alters the time profile of liquid argon scintillation.
The time structure of pure liquid argon scintillation light contains fast ($\sim 6$~ns) and slow ($\sim1.6~\mu \mathrm{s}$) components that correspond to the decays of singlet and triplet states of the Ar$_2$ dimer, respectively \cite{Kubota1978}. The relative quantities of the singlet and triplet states depend strongly on the initial ionization density at the location of the deposited energy and consequently enable the identification of the initial interacting particle \cite{Hitachi1983}. The presence of xenon adds several new pathways for energy transfer that produce ArXe and Xe$_2$ dimers \cite{Galbiati:2021,Fields2022},
both by reacting with Ar$_2$ dimers before they decay and by reacting with their excited atomic argon precursors \cite{Gan2022,Galbiati:2021,Fields2022}.  This alters the time structure of the scintillation light in addition to channeling energy toward the longer wavelengths.  Improvements from xenon doping in the ability of argon to distinguish between events resulting from electronic and nuclear recoils through the time structure of primary scintillation light are being debated, with reports of both beneficial and detrimental effects \cite{Pfeiffer2008,Wahl2014}.  Adding xenon speeds the emission of scintillation light at all concentrations \cite{Kubota1993TailSuppression}.  The sharper signal allows for reduced waveform acquisition times and reduces the chance of pileup or accidental coincidence, allowing for new applications as an inexpensive fast scintillator \cite{Vogl2022,Ramirez2021,Lai2021}. These benefits have lead to the consideration of xenon doping by the LEGEND experiment~\cite{Mcfadden2021}, the Scintillating Bubble Chamber project~\cite{SBC2022} and large liquid argon TPC neutrino experiments including protoDUNE~\cite{Whittington2019,Gallice2022}. 

Xenon doping above 1000~ppm offers additional benefits to ionization signal detection. First, the presence of xenon in liquid argon leads to 
an increase (10--15\%) in charge yield that is attributed to the Penning ionization of xenon by Ar$_2$ dimers \cite{Suzuki1993LightOA,Kubota1976,Hagmann2004}.  Second, percent-level xenon doping in liquid argon results in the presence of tens-of-ppm xenon in the gas phase of a dual-phase argon detector.  This may increase the electroluminescence photon yield per drifted electron. 
In these detectors, electroluminescence is produced by the inelastic collision of gas phase argon atoms with electrons moving towards the anode after they are extracted from the liquid surface under high field. 
When xenon is present in the gas, due to its lower excitation energy relative to that of argon, electrons will transfer energy to it disproportionately and produce more primary excitation.  
Finally, significant xenon presence in the gas can substantially improve the efficiency of detecting electroluminescent ionization signals in dual-phase TPCs. The addition of xenon in the gas phase modifies the chemical reactions following excitation formation and can lead to wavelength shifting in a way closely analogous to the processes in the liquid \cite{Kubota1976}. Although the electroluminescent benefits of xenon doping in a dual-phase argon TPC have yet to be demonstrated, a few similar experiments allow prediction of the behavior.  A proportional chamber operated at room temperature and atmospheric pressure found that doping argon with 77~ppm xenon was sufficient to transfer most of the energy from the 128~nm peak to a 147~nm peak attributed to ArXe.  Doping to 1013~ppm resulted in transfer of most of the energy to a broad dominant peak at 171~nm, attributed to Xe$_2$ \cite{Takahashi1983}.  Another experiment exposed a range of mixtures at pressures varying from 400 to 1400 mbar at room temperature to a beam of 640~MeV argon ions.  The resulting light is very similar to that of the proportional counter and a doping of 30~ppm was sufficient to replace the majority of the 128~nm emission with longer wavelengths \cite{Efthimiopoulos1997}.  
The spectra of gas phase doped mixtures at room temperature and atmospheric pressure differ from those of liquid phase mixtures in that there is a substantial xenon concentration range around 50~ppm in which the predominant light emission is from the ArXe dimer.  Conversely, in doped liquid this line is quenched by further transfer of energy to the Xe$_2$ dimer at similar doping ratios \cite{Neumeier2015A}. 
Although the lifetime of the ArXe dimer has not been measured, it is expected to be much shorter than that of Ar$_2$~($\sim3~\mu$s), and this will improve the detection of low-energy ionization signals and separating events.

Therefore, a heavily doped dual-phase argon detector that efficiently detects wavelengths of 147~nm and longer with VUV SiPM sensors~\cite{HamamatsuVUVGraph} can operate without wavelength shifting coatings and achieve improved sensitivities to low-energy ionization signals. 
Compared to dual-phase xenon TPCs, an argon detector can have lower background electron emission rates due to the suppressed impurity outgassing at a lower temperature and the higher efficiency of extracting electrons from liquid into the gas. 
Such an argon detector would be especially advantageous for detecting interactions with low mass particles such as neutrinos and light dark matter due to their stronger kinematic coupling to argon than that to heavier targets. 
The concept of using heavily xenon-doped liquid argon as a target for neutrinoless double beta decay searches was also proposed recently~\cite{Mastbaum:2022}. 

Xenon doping technology needs to be systematically studied to realize the benefits of xenon-doped argon.  To date, the thermodynamic stability of xenon doped argon detector systems has not been thoroughly investigated.  Various forms of fast and slow instabilities can develop in xenon-doped argon systems~\cite{Wahl2014,Ishida1997,XinranLi2020}. 
In this paper, we discuss the different modes of argon-xenon mixture instabilities in Sec.~\ref{sec:instability}, 
and explain how the apparatus used in this work is designed to mitigate these at percent-level xenon concentrations (Sec.~\ref{sec:apparatus}). Section~\ref{sec:measurement} demonstrates the performance of this system and confirms effectiveness of the mitigation strategies; Section~\ref{sec:concl} concludes this work.

\section{Instability of Argon-Xenon Mixtures}
\label{sec:instability}

Xenon dissolves efficiently in liquid argon up to a solubility limit of 4--8\% that increases with temperature~\cite{Yunker1960}.
However, instabilities can develop even in mixtures with concentrations substantially below the solubility limit~\cite{Ishida1997,Wahl2014}, causing the xenon concentration to increase in one part of the system and decrease in others. 
In extreme cases, xenon may precipitate from the liquid and form ice on the surfaces of solid detector components.  This can substantially deplete the mixture of xenon and also has mechanical consequences such as the blockage of fluid flow~\cite{Wahl2014, XinranLi2020}. 

The origin of instabilities in xenon-doped argon lies in the vastly different vapor pressures of xenon and argon at a given temperature. In a low-pressure system at 1~bar liquid argon can only exist in a narrow temperature range of 83.8--87.2~K. 
At these temperatures, pure xenon exists in the solid form with a small vapor pressure of $\sim$40 ppm relative to argon.  
When xenon is dissolved in liquid argon, 
its vapor pressure above the liquid mixture is expected to be no higher than that of the solid vapor pressure at any xenon concentration in the liquid. 
Consequently, the relative xenon mole fraction in the vapor is greatly suppressed from that in the liquid. 
Details of the xenon partial pressure are given in the Appendix~\ref{subsec:solubility}.

\begin{figure}[!ht]
\centering
\includegraphics[width=0.49\textwidth]{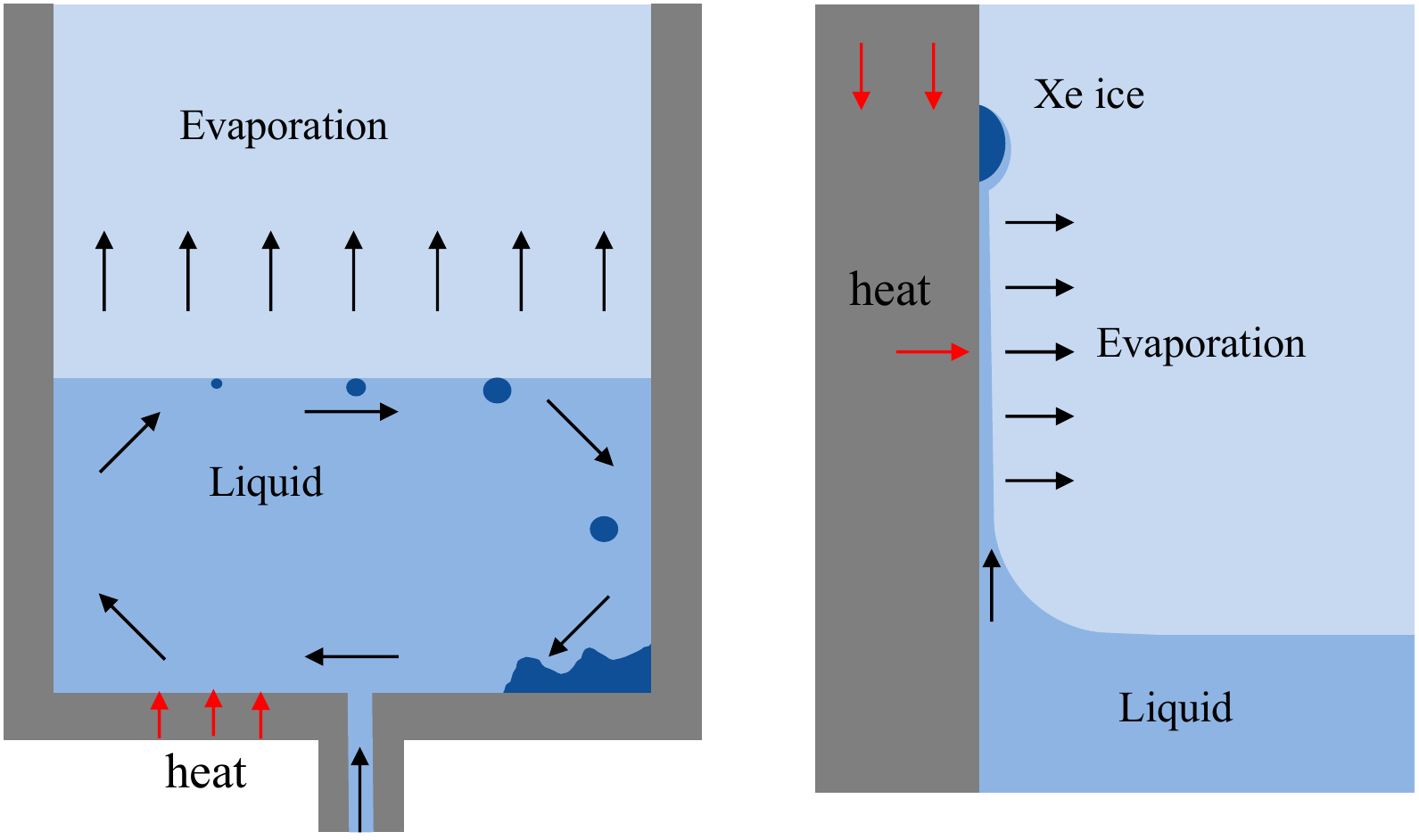}
\caption{Conceptual illustrations of the distillation instabilities in xenon-doped liquid argon. {\bf Left:} the evaporation of a liquid argon-xenon mixture produces vapor with reduced xenon concentration and liquid with enhanced xenon concentration, which can lead to formation of xenon crystals when the solubility limit is exceeded; {\bf Right:}  surface tension draws a thin film of liquid onto the detector wall that is isolated from the main bath.  Evaporation due to heat provided by the wall concentrates xenon in the film and leads to xenon ice formation.}
\label{fig:distillation}
\end{figure}

The large disparity of xenon and argon vapor pressures can cause subtle but serious problems for xenon-doped liquid argon detectors.  
The first and most prominent problem relates to how a typical liquid argon detector removes impurities from the liquid to maintain a high purity for its operation. 
This is conventionally achieved by withdrawing liquid, evaporating it, purifying the resulting gas, condensing that gas, and finally returning the liquid to the detector volume.  A system may evaporate and condense the argon on opposite sides of a heat exchanger to reduce cooling requirements, and may also use a counterflow gas phase heat exchanger for further efficiency.  
As illustrated in Fig.~\ref{fig:distillation} (left), when xenon is present in the liquid, the evaporated gas leaving the liquid surface will contain a much lower concentration of xenon than that in the liquid due to the low vapor pressure of xenon. As this process continues, xenon becomes concentrated near the evaporating surface while the extracted gas to be purified and recondensed is nearly depleted of xenon.
This scenario is analogous to a single stage of chemical distillation. 
When the xenon concentration at the evaporation surface exceeds the solubility limit in such a system ice will form and its accumulation may block the circulation path of the detector~\cite{XinranLi2020}.
This problem may be avoided by purifying the liquid mixture directly--an approach being pursued for large argon and xenon systems~\cite{Aprile2022_LiquidPurification}-- rather than removing impurities from the gas phase. 

The second mode of instability in liquid argon-xenon mixtures is analogous to the first one but more subtle. Even in a xenon-doped liquid argon detector that does not deliberately evaporate liquid, unintentional evaporation at liquid surfaces may still occur. As shown in Fig.~\ref{fig:distillation} (right), detector components that touch the liquid surface, such as the container vessel of the liquid and electrical cables, can transfer heat to the liquid, as can submerged electronics. 
Such heat will result in an unintended evaporation of the liquid and enrich xenon in the vicinity of the point of evaporation.  If this highly concentrated liquid cannot mix with the unsaturated liquid bulk it may eventually lead to xenon precipitation near these evaporation locations and xenon depletion from the main liquid volume.
This effect is augmented by the surface tension of the liquid, which can transfer a small amount of liquid away from the main bath while increasing the contact surface with heating sources, as demonstrated by the xenon ice buildup to be explained later in this work.  Modeling this instability is complex because it depends on the detailed temperature and pressure profile of the whole system, the heat flows at points of liquid contact and the flow pattern within the liquid. 

A third source of instability in an argon-xenon mixture is the introduction of xenon-rich gas into the liquid argon volume.  
A typical liquid argon cryogenic system condenses warm gas by exposing it to a cold surface; after the gas condenses it flows down to the detector volume by gravity. 
Given the low vapor pressure of xenon at liquid argon temperature, such a condensation scheme can only function if the argon gas contains no more than tens of ppm of xenon. When the xenon concentration exceeds the saturation vapor pressure excess xenon will solidify on the cold surface encountered by the gas before it reaches the liquid~\cite{Wahl2014}. 
This phenomenon limits the rate of xenon introduction into a liquid argon system, especially one that requires a relatively large quantity of xenon either because of a large overall volume or a high doping concentration.
As explained in Sec.~\ref{subsec:cryosystem}, this problem can be mitigated by introducing xenon-rich gas directly into the liquid argon volume, but the proper implementation requires the cryogenic system to be significantly different from that of a conventional liquid argon detector.

\section{Experimental Setup}
\label{sec:apparatus}

The experiments were carried out using the CHILLAX (CoHerent Ionization Limit in Liquid Argon and Xenon) test stand, which was specifically designed to address the challenges of high-concentration xenon doping in liquid argon and to study its benefits for both scintillation and ionization signal detection. 
The test stand features a unique cryogenic system that enables efficient condensation of xenon-rich argon at the percent level and stabilizes the liquid argon-xenon mixture under a range of operating conditions. 
The system is also equipped with diagnostic tools to monitor the xenon concentrations in the liquid and in the gas and the accumulation of xenon ice at different parts of the cryogenic system. 

\subsection{Cryogenic system}
\label{subsec:cryosystem}

\begin{figure}%
\centering
\includegraphics[width=0.45\textwidth]{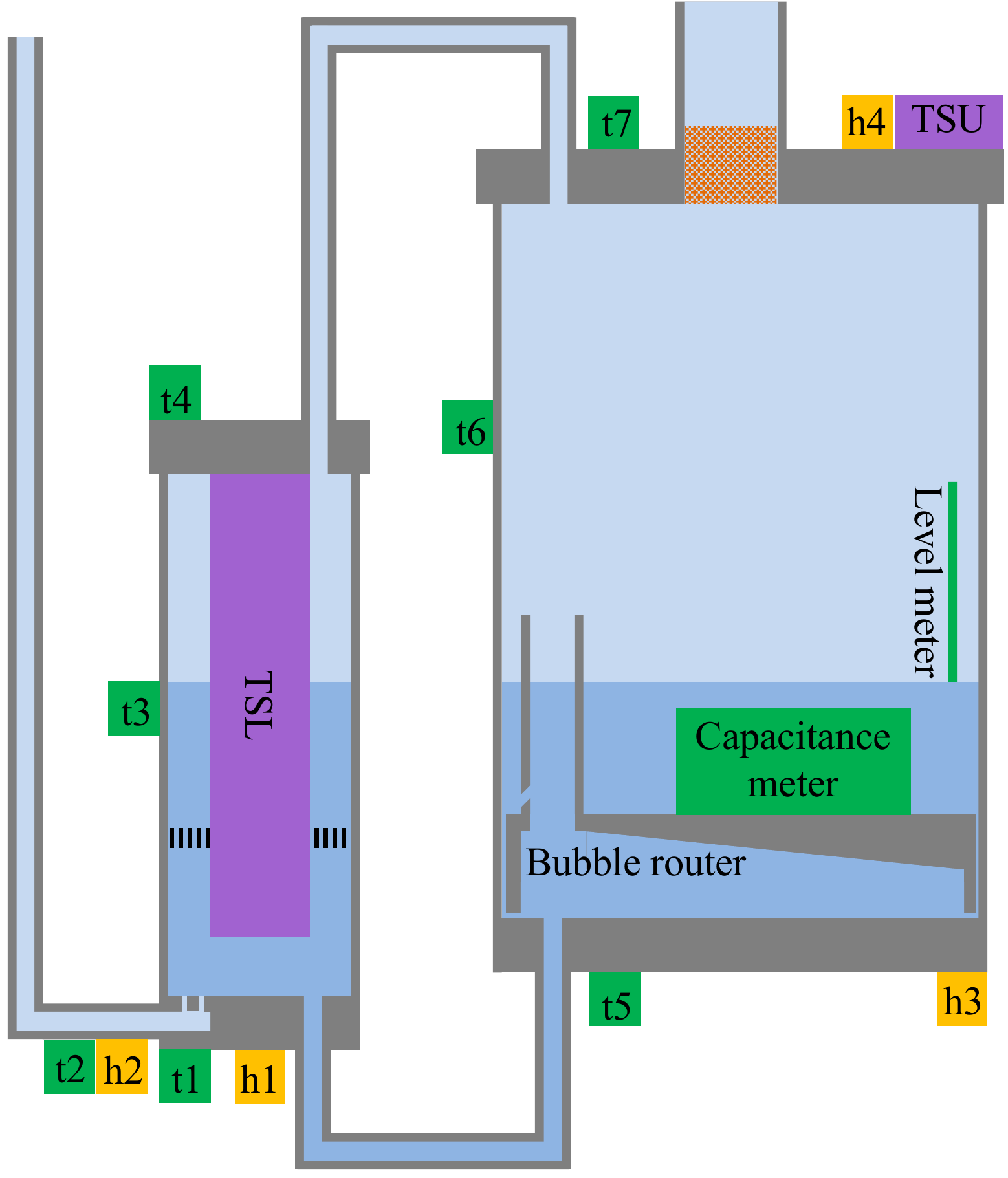}
\caption{A block diagram of the CHILLAX cryogenic argon handling system with the heat exchanger (HX) shown on the left and the detector on the right; the two volumes are connected both in the liquid and gas phases. TSU and TSL indicate the upper and lower thermosiphon evaporators, respectively; t1--t7 are thermometers and h1--h4 are heaters. The capacitor meter, bubble router, liquid level sensor and the liquid delivery line into the HX are also shown. }
\label{fig:chillaxpid}%
\end{figure}

The liquid argon and xenon mixture is held in a 4.5 inch diameter stainless steel can with a 6.75 inch CF flange that integrates electrical, gas, and optical ports. 
The detector flange is supported by a room-temperature cryostat flange through a 1 inch diameter stainless steel tube that also serves as a pathway for gas and cabling. 
The detector enclosure is wrapped with multi-layer aluminized mylar film and situated inside a customized vacuum cryostat to provide thermal insulation. 
The cryogenic system is powered by a SHI CH104 cryocooler ($\sim$50~W at 85--95~K) that cools two independent thermosiphon loops.
The working fluids (argon is used in this experiment) of the two thermosiphons are completely separate from the detector volume containing the argon-xenon mixture.
As illustrated in Fig.~\ref{fig:chillaxpid}, the upper thermosiphon loop (referred to as the TSU), provides cold liquid argon to a copper evaporator mounted on the detector flange where an electric heater is attached to regulate the flange temperature.  
The lower thermosiphon loop (TSL) cools a dual-phase heat exchanger (HX) installed near the bottom of the detector volume that is responsible for gas condensation. 
The TSL liquid argon flows to the inner section of the HX by gravity and cools the outer annular volume of the HX, where argon gas with or without xenon condenses. 
The temperature and pressure of the argon in the TSL are controlled by heaters through a PID feedback loop.  

Gas delivered to the bottom of the outer HX volume condenses in the annular volume; the liquid then flows into the detector through a line connecting the bottoms of the two volumes.  The top of the outer HX volume is also connected to the top of the detector to form a pressure interlock that equalizes the head pressures and thus liquid levels of these spaces. 
The detector pressure and the corresponding liquid temperature are controlled primarily by the selection of the temperature of the TSL liquid, with fine tuning achieved using a second PID loop that regulates a heater mounted on the bottom of the detector can. 
In steady-state operations, the feedback loop is capable of limiting pressure variations to within $\pm$1~mbar of the set point with an average heating power of $\sim$0.5~W delivered to the detector bottom.
The system was designed to operate at pressures up to 2.5~bar absolute; both the xenon solubility and the xenon vapor pressure increase with pressure, which provides benefits to highly doped systems and dual-phase operation. 

To stabilize the liquid surface in presence of heating from the PID feedback and from radiation heat, the detector is fitted with a bubble router that steers bubbles generated at the bottom of the detector away the liquid surface.  The router is shaped like an inverted cup with a gradual slope on the underside to collect rising bubbles and release them through a vertical tube ending above the liquid surface.  A thin (1~mm) upwardly sloped slot in the fitting connecting the lower end of the tube to the bubble router provides a liquid connection between the volumes above and below the router without diverting rising bubbles from the tube. 

The experimental setup is also equipped with an external circulation loop that can purify the gas and add xenon to the detector.
A metal bellows compressor (Senior Bellows) extracts gas from the top of the main detector bath through the 1~inch support tube, pumps it through a hot zirconium getter (SAES Monotorr), and returns the purified gas to the outer HX volume to be recondensed. 
To add xenon to the liquid argon, a user-specified flow rate of xenon is injected into the circulation path where it mixes with the circulation argon gas before being purified and condensed in the HX. 
To avoid xenon condensing or freezing before it enters the HX, a tube-mounted heater near the gas entrance to the bottom of the HX warms the incoming gas stream to over 200~K during xenon addition. After xenon enters the HX, it condenses and mixes with the liquid argon.  
A fine horizontal stainless steel wire screen is positioned in the HX liquid to trap large bubbles of the injected xenon-rich gas mixture from rapidly rising through the liquid argon inside the HX and condensing on the cold surfaces of the TSL boundary. 
The rising introduced gas and the thermal expansion of the liquid at the gas entrance produce an upward moving convective flow that rapidly mixes the liquid inside the HX annular space.   

During steady-state circulation without xenon doping, the distillation effect retains the high concentration of xenon in the detector liquid bath. 
Xenon uniformity in the main bath is provided by natural convection due to evaporative cooling at the surface and modest heating at the bottom. 
The concentration of xenon in the vapor phase is greatly reduced relative to that of the liquid phase by the large Henry's law constant $H\sim ~600$ (Sec.~\ref{subsec:solubility}). Consequently, the gas streams provided to the HX through the circulation pump and the pressure interlock contain very little xenon.  
The low concentration of xenon in the HX liquid argon is maintained because the continuous flow of liquid from the HX towards the detector prevents diffusion of xenon from the detector into the HX. The transport of xenon by the liquid flow exceeds its transport by diffusion by a factor (the P\'eclet number) of $\sim~4000$ when calculated for our slowest flow rates (300~SCCM) under the conservative assumption that xenon diffuses no faster than argon \cite{ArLiqSelfDiff}.   
This low xenon concentration also serves to prevent the formation of xenon ice on the TSL boundary surfaces, which are the coldest xenon-exposed surfaces of the system, and any ice that does form during the doping phase is washed over and dissolved by xenon-depleted argon. 

\subsection{Xenon concentration measurement in the liquid}
\label{subsec:capacitancemeter}

Liquid argon has a dielectric constant of 1.505 at 1~bar pressure on its vapor curve. The value for liquid xenon is 1.85 at the same condition.  Consequently, the dielectric constant of a liquid argon-xenon mixture varies with the relative xenon concentration, and
the capacitance between two well-positioned electrodes can provide a measurement of the xenon doping level. 
We installed a custom parallel plate capacitor 1.14 inches below the liquid level in the detector cryostat (Fig.~\ref{fig:chillaxpid}) to measure the xenon concentration within the mixture. 
We assume the capacitance to be linear with the dielectric constant of the fluid between the electrodes, and the dielectric constant is approximately a density-weighted average of the dielectric constants of its constituents at low doping concentrations. More details are given in Section~\ref{subsec:dielectricDependence}.
The calibration factor between capacitance and xenon concentration was obtained from data acquired during the doping process, which is elaborated on in Section~\ref{subsec:doping}. 

The capacitor assembly includes a pair (termed 'active' and 'reference') of parallel plate capacitors that share a common ground electrode.  They are identical except that the space between the electrodes of the active capacitor is open to the liquid mixture, while this space in the reference capacitor is displaced by PTFE, minimizing its sensitivity to xenon doping. 
Their readout components are closely matched to achieve correlated systematic errors.  The measured capacitance in this work is always that of the active capacitor minus the reference capacitor. 

Capacitance is measured with a Texas Instruments 4-channel FDC1004 chip. The FDC1004 measures capacitance by sending a 25 kHz voltage stepped waveform to the (active or reference) sense electrodes and measures the corresponding charge transfer with an 
analog to digital converter (ADC).  
RG178 coaxial cables connect the ground and two sensing electrodes to the FDC1004 chip through floating-shield coaxial feed-throughs.  The FDC1004 chip also supplies an active shield voltage to the outer conductor of the coaxial cables that excludes the capacitance of the cables from the measurement.
This method achieved a capacitance sensitivity of $<1$~fF after averaging, which corresponds to a xenon concentration in liquid argon of 0.05\%. 

\subsection{Monitoring Instrumentation}
\label{subsec:inst}

The CHILLAX system is also equipped with a camera to provide a live view of the detector interior and a gas sampling system to measure the xenon concentration in the gas.

To provide a visual indication of the thermodynamic state, 
the detector bath is observed from above by a digital camera (Raspberry Pi camera module V2) situated above a 2.75 inch CF sapphire viewport on the detector flange. The camera is thermally isolated from the detector flange and is electrically heated to $\sim 0$ C.
Lighting in the detector volume is provided by two LEDs, one each pointed upward and downward.
A stainless steel dental mirror is mounted above the liquid surface to allow a partial camera view of the underside of the detector flange. Image data from the camera is transferred from the vacuum cryostat through a standard 15 pin D sub connector adapted by custom PCBs to the camera's flat flexible cable. 
This design was inspired by the camera system of the ProtoDUNE-SP experiment \cite{Kordosky2017}. 

The xenon concentration in the gas phase is monitored by a custom-built sampling system with a Stanford Research Systems Residual Gas Analyzer (RGA). The sampling system can sample from three locations in the circulation path: the gas outlet from the detector, the argon-xenon mixing volume, and the outlet of the getter. Gas samples are first reduced to a pressure of approximately 2~torr through volume expansion into a 1~L sampling cylinder, which also serves as a reservoir for the sampling system. The gas pressure is reduced further by a factor of $10^{5}$ by pumping through a $75~\mu$m orifice produced by Lenox Laser (SS-4-VCR-2-75) before it is sampled by the RGA. 
    
Calibrations of the sampling system are performed by creating known mixtures of xenon in argon in the 1~L cylinder and then sampling them with an initial pressure of 2~torr to mimic the sampling conditions of CHILLAX. The calibration mixtures are created by adding xenon at controlled pressures 
to the 1~L reservoir, which is then diluted with the addition of pure argon at controlled pressures. 
Fine control of the xenon and argon pressures is achieved with volume expansion and accurate measurements from an MKS 626C capacitance manometer (100~torr range) and Setra 225 pressure transducer (3.4~bar range).

Gas samples from the detector outlet were regularly taken and measured with the RGA system. 
At $\sim$2\% xenon concentration in the liquid argon, we measured approximately 30--50~ppm of xenon in the gas by extrapolating from over a dozen prior calibration samples spanning 5--50~ppm. 
Due to a not-yet-understood issue with the sampling system that caused the measured xenon concentration to drift over time, we conservatively report the lowest reading obtained.
This result is consistent with the Henry's law calculation as explained in Sec.~\ref{subsec:solubility}. 
As discussed in Sec.~\ref{sec:intro}, 30--50~ppm of xenon in the vapor is expected to produce significant benefits for the detection of ionization signals in a dual-phase TPC. 
Therefore, we anticipate a 2\% argon-xenon mixture to be adequate for a dual-phase TPC exploring low-energy ionization signal detection.

\section{Measurement and Results}
\label{sec:measurement}

We performed a series of experiments to demonstrate that the CHILLAX system is capable of promptly incorporating xenon-rich argon gas into its liquid argon target medium, and that it can stabilize and actively circulate mixtures of 2.35\% xenon-in-argon for several days without degradation.
For contrast, we also operated the detector in modes where the thermal environment of the detector was allowed to deviate from optimum states
and observed segregation of the xenon from the liquid argon through xenon ice growth on specific detector surfaces.
These observations confirm our understanding of the thermodynamic behavior of xenon-doped argon and the proposed approach to mitigate the instability of such doped systems. 
Implications of this work for future xenon doping efforts are also discussed. 

\subsection{Introduction of xenon into liquid argon}
\label{subsec:doping}

The experiment began with the condensation of 454 standard liters of pure argon gas in the heat exchanger (HX). The argon liquid was continuously delivered to the detector volume and formed a liquid bath of 6.6~cm height. The liquid level in this work is informed by two closely spaced platinum thermistors that self-heat to different temperatures depending on the thermal conductivity of their immediate surroundings. 
This implementation is used to mitigate interference from xenon doping in a capacitance-based level meter. 
The condensation was stopped when the liquid surface was sensed by the level meter, and this height was visually confirmed with the camera.
Next, a gas circulation rate of 1.5~SLPM was established, with gas removed from beneath the detector flange and returned to the HX for recondensation. 
The external portion of the circulation path is plumbed for purification of the gas, 
but the purifier was bypassed for this experiment because a high chemical purity is not required for cryogenic studies.

The xenon concentration of the liquid mixture was measured through changes in its dielectric constant, as described in Section~\ref{subsec:capacitancemeter}. 
Early experiments showed the capacitance meter is also sensitive to density changes caused by sub-Kelvin temperature fluctuations of the mixture.  For this reason, the doping was performed only once the detector had thermally stabilized using the method 
explained in Sec.~\ref{subsec:cryosystem}. 
A stable detector pressure at 1.8$\pm$0.005 bar with a liquid batch temperature of 93.3~K was achieved through PID control of the heating power (averaging 0.5~W) applied to the outside of the bottom flange of the detector can.  

\begin{figure}[!t]
\centering
\includegraphics[width=0.99\linewidth]{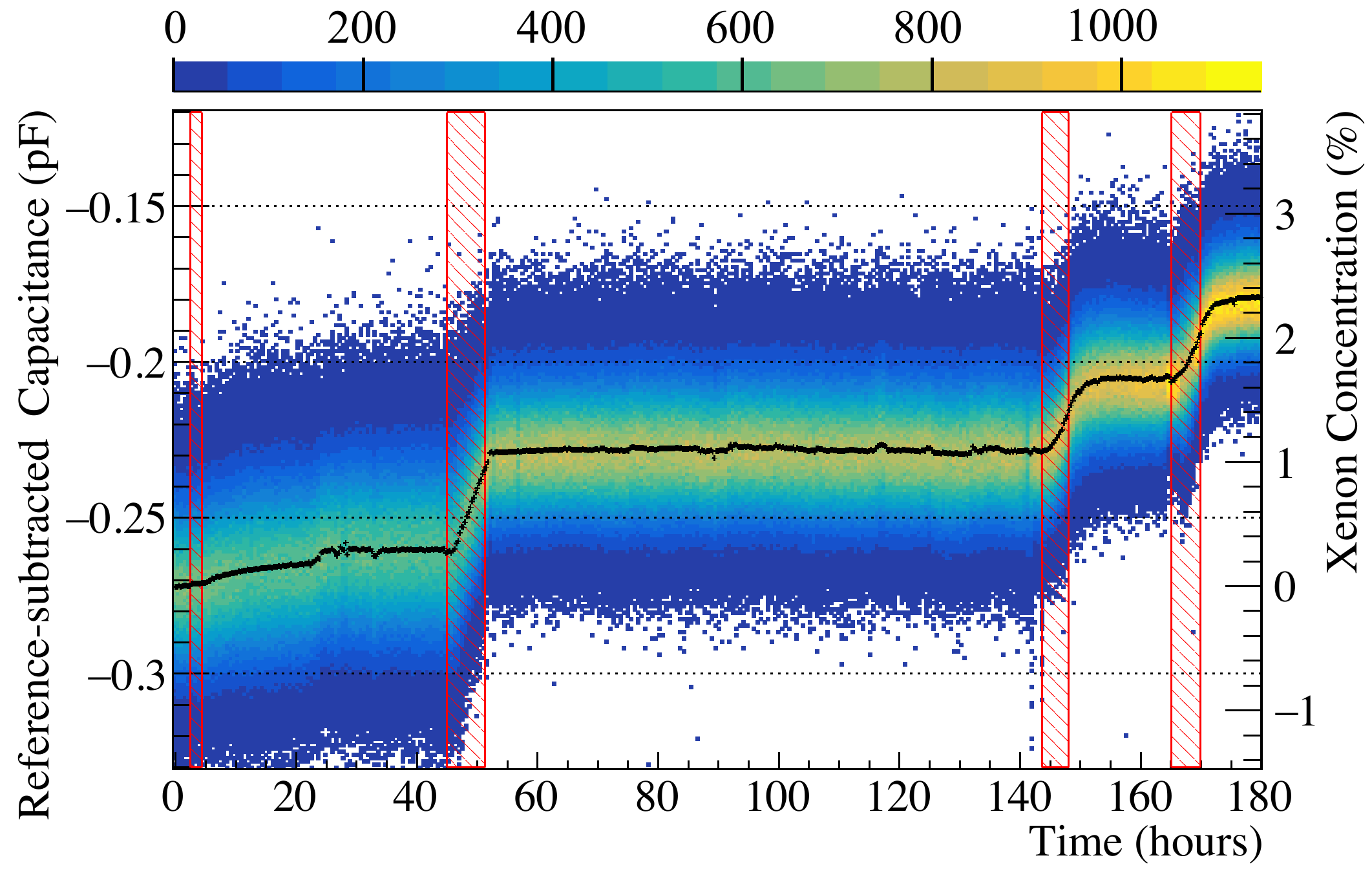}
\caption{Changes of measured capacitance (left y-axis) and estimated xenon concentration in liquid argon (right y-axis) during the doping process. The heat map represents the raw capacitance data with the mean after averaging shown in black; red vertical bands indicate the four doping periods.}
\label{fig:doping}
\end{figure}

Xenon was introduced into the liquid argon volume in four separate steps.
In each step, 8.8~sccm of xenon gas was injected into the mixing volume in the circulation path. 
This corresponded to 0.6\% xenon in the overall gas flow of 1500~sccm, which is two orders of magnitude higher than the relative vapor pressure of xenon at liquid argon temperature.  
Prior to and during doping, the inlet tube to the HX is heated to over 200~K to prevent xenon from condensing or freezing before entering the HX. The heating also enhances liquid convection and mixing in the HX.
For the four doping steps, the xenon flow continued for 114, 377, 266 and 289 minutes, increasing the xenon concentration in liquid argon by 0.26\%, 0.86\%, 0.59\%, and 0.64\%, respectively. 

As shown in Fig.~\ref{fig:doping}, with the exception of the first doping step, the measured capacitance increased approximately linearly during the doping period and stabilized soon after xenon introduction was stopped. During the first doping, the detector flange temperature was set below that of the detector liquid bath and that of the HX, 
causing argon gas to condense on the flange and flow down the detector walls.  
This was seen visually through the camera and independently confirmed by the temperature of the detector wall measured a few inches above the liquid surface.  Under normal conditions without argon film flowing down the wall, the temperature of the wall thermometer is typically 1--2 K above that of the detector bath due to the imperfect isolation from infrared radiation.  In contrast, when argon liquid film flows down the detector walls it tightly anchors the wall temperature to the vapor curve, close to that of the bath.  The condensation on the detector flange likely caused the gas from the HX to flow into the detector through the gas interlock line.
As a result, the flow of xenon-rich liquid from the HX to the main bath was greatly diminished and the xenon concentration in the main bath rose very sluggishly during and after the first doping step.  
In subsequent doping steps, we set the detector flange temperature above that of the HX and detector bath to encourage gas flow from the detector into the HX.  Xenon introduced into the HX in this state appeared promptly in the main detector bath. 

After doping ceases we anticipate nearly all of the residual xenon in the HX to eventually migrate into the detector volume.  This is because both the internal and external circulation loops feeding the HX are supplied with gas evaporated from the detector main bath surface, which is reduced in relative xenon concentration by a factor of $H$ ($\sim$600, see Sec.~\ref{subsec:solubility}) compared to the liquid in the bath.  
When this xenon-depleted gas is condensed in the HX, it dilutes the xenon concentration in the HX; the constant flow of liquid from the HX toward the detector bath transfers the residual xenon into the detector and also prevents diffusion of xenon from the detector bath back into the HX. 
This is confirmed by the observation of a continued increase of xenon concentration in the detector bath up to a few hours after the last three doping steps, followed by a plateau; 
once the plateaus are reached only a negligible amount of xenon ($1/H$ in concentration relative to the detector bath) is located in the HX.
During the second doping, the temperature at the top detector flange accidentally fluctuated below the vapor curve (while remaining above the HX temperature) and caused argon gas to condense; this tended to reduce the detector pressure and triggered the PID loop to increase power substantially to the bottom heater to maintain the system pressure at 1.8~bar. 
As a result, a stronger gas flow from the detector to the HX augmented the usual liquid flow from the HX to the detector volume and produced a faster transfer of xenon into the detector. 
For the last two doping tests, the detector flange temperature fluctuation issue was resolved and the time scale of the approach to the plateau was consistent with the nominal liquid flow rate from the HX to the detector driven by the external gas circulation rate.

Although nearly all the doped xenon was delivered to the detector, the exact xenon distribution within the detector volume is complicated by the bubble router plate (Sec.~\ref{subsec:cryosystem}) that divides the detector bath.  
Above it is the super-router region containing the bath surface and concentration meter; below it is the sub-router region in contact with the bottom of the detector can where liquid arrives from the HX.  The sub-router volume is 14\% of the total detector bath volume.  In principle, the bubble routing plate can lead to unequal distribution of xenon in these two volumes, 
as a unidirectional liquid flow through a small passage would oppose xenon diffusion.  However, the lower lip of the bubble routing plate is not sealed to the bottom flange of the detector can, and bubble production beneath the router plate and their passage through the bubble routing tube can cause a breathing motion that drives an oscillatory exchange of liquid between the two volumes.

\begin{figure}[!ht]
\centering
\includegraphics[width=0.45\textwidth]{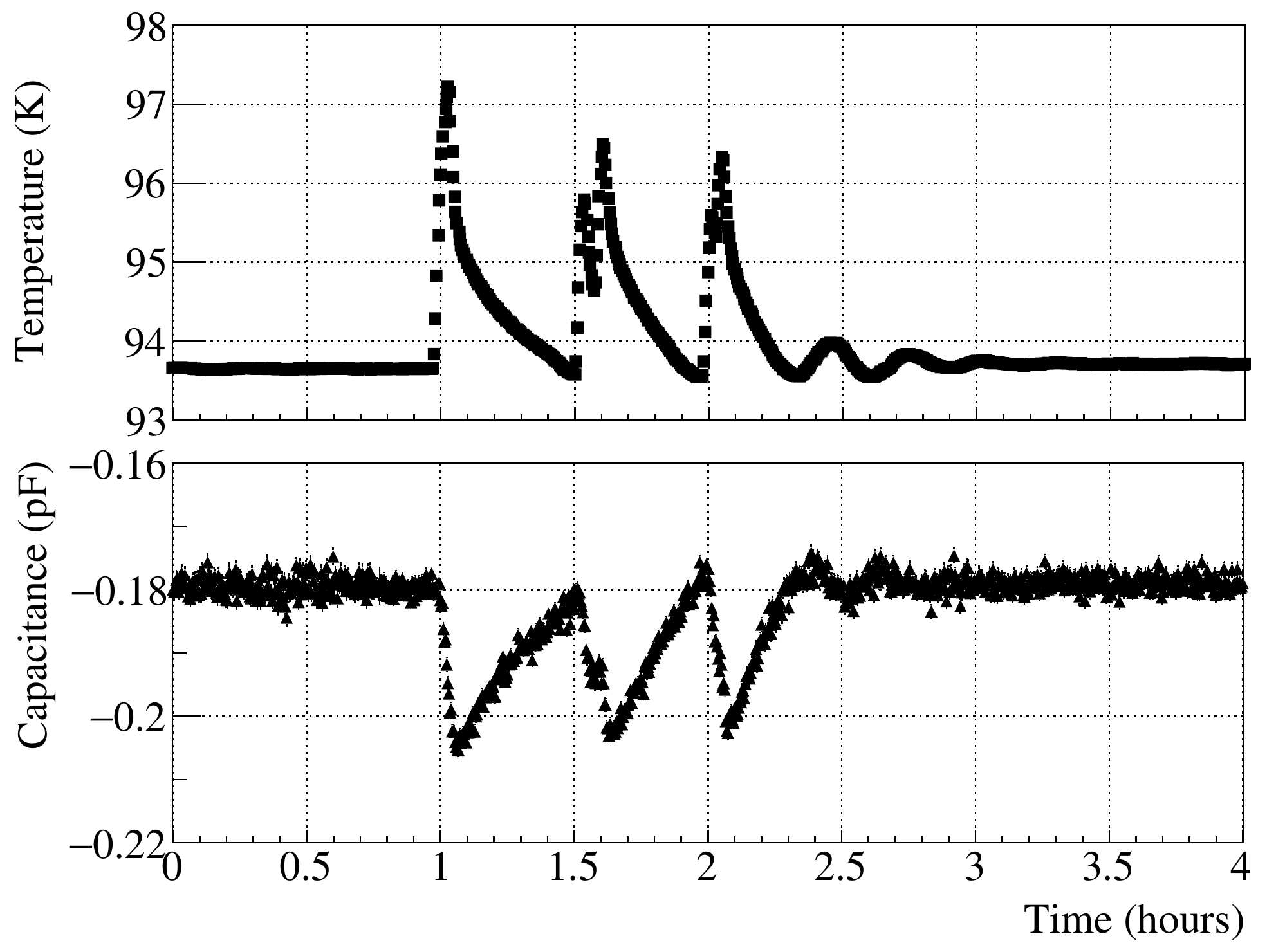}
\caption{The temperature of the liquid mixture (top)  and the reference-subtracted capacitance (bottom) measured during the detector volume mixing tests. The decreases of capacitance are due to expansion of liquid at elevated temperatures rather than xenon concentration changes. }
\label{fig:uniform}
\end{figure}

To estimate the difference in xenon concentration between the super-router and sub-router regions we conducted a series of tests to enhance liquid mixing.  The tests followed a long period (overnight) of steady operation at 2.35\% doping in which a concentration difference could have developed. Then rapid mixing was driven by applying a large heating power of 30~W to the detector bottom flange for several minutes while maintaining a safe detector pressure.  This created vigorous bubbling both through the bubble routing tube and around the lower lip of the bubble router, enhancing liquid exchange between the two volumes.  
We presume that a large fraction of the liquid supplied to the sub-router region arrived from the super-router region by passage around the lower lip of the router.  

This process was repeated three times in rapid succession; between subsequent tests we allowed about 30~minutes for the detector temperature and pressure to return to the standard measurement condition.
As illustrated in Fig.~\ref{fig:uniform}, for all three tests, the measured capacitance values returned to the premixed ones within measurement uncertainty, suggesting that any xenon concentration difference between the regions above and below the bubble router is negligible.  
Note that the drop of measured capacitance during these tests was a result of increased temperature (via decreased liquid density) rather than a change of xenon concentration. 
Therefore, entire detector bath volume is used for a calculation of xenon concentration (right y-axis in Fig.~\ref{fig:doping}) and for the calibration of the capacitance meter.
If insufficient liquid mixing between the sub-router and super-router regions allowed the sub-router region to become xenon depleted, our calculation would underestimate the xenon concentration in the main detector bath by as much as 14\%, so our reported results should be seen as conservative. 

\subsection{Stability of the Xe-Ar mixture}
\label{subsec:stabilitytest}

A xenon concentration of over 2.3\% was present in the liquid argon volume after the completion of the xenon doping, which is only a factor of 3 below the solubility limit of xenon in liquid argon at the operating temperature. 
This high xenon concentration, together with the continuous circulation of the detector gas, augments the instability of xenon-doped liquid argon. As described later in this section, when the thermal profile of the system is not carefully controlled, xenon can concentrate near the liquid surface, deposit as a layer of ice on detector surfaces, and segregate from the liquid mixture. We also demonstrate that with proper control of the temperature field the liquid mixture can be stabilized for at least multiple days with no degradation.

In the first test, we created a uniform temperature field in the system to reduce undesired heat flow by
maintaining a stable detector flange temperature of 0.3--0.5~K above that of the liquid. 
This slightly higher temperature prevents argon gas from condensing on the detector flange while introducing minimal heat flow from the flange to the liquid through the conduction of the walls.
Additional steps were taken to improve the homogeneity of the temperature field across the whole system. 
Upon completion of the last doping step we gradually stopped heating the tubing at the inlet of the HX condensing volume and reduced the external gas circulation rate to 300~SCCM~\footnote{The circulation speed was found to not significantly disrupt the mixture stability in CHILLAX. This is demonstrated by the capacitance data between the second and third doping steps, when the circulation speed was 1500~SCCM (Fig.~\ref{fig:doping}).}.  
To offset this heat reduction at the HX, the temperature of TSL was increased accordingly to maintain a stable detector pressure of 1.8~bar and a low average heating power on the detector bottom of about 0.5~W.  With these changes, the HX temperature became significantly lower than that of the detector bath and an unambiguous internal gas circulation pattern was established in which gas flowed continuously through the pressure interlock tube toward the HX, while liquid flowed continuously from the HX to the detector.

\begin{figure}[!t]
\centering
\includegraphics[width=0.99\linewidth]{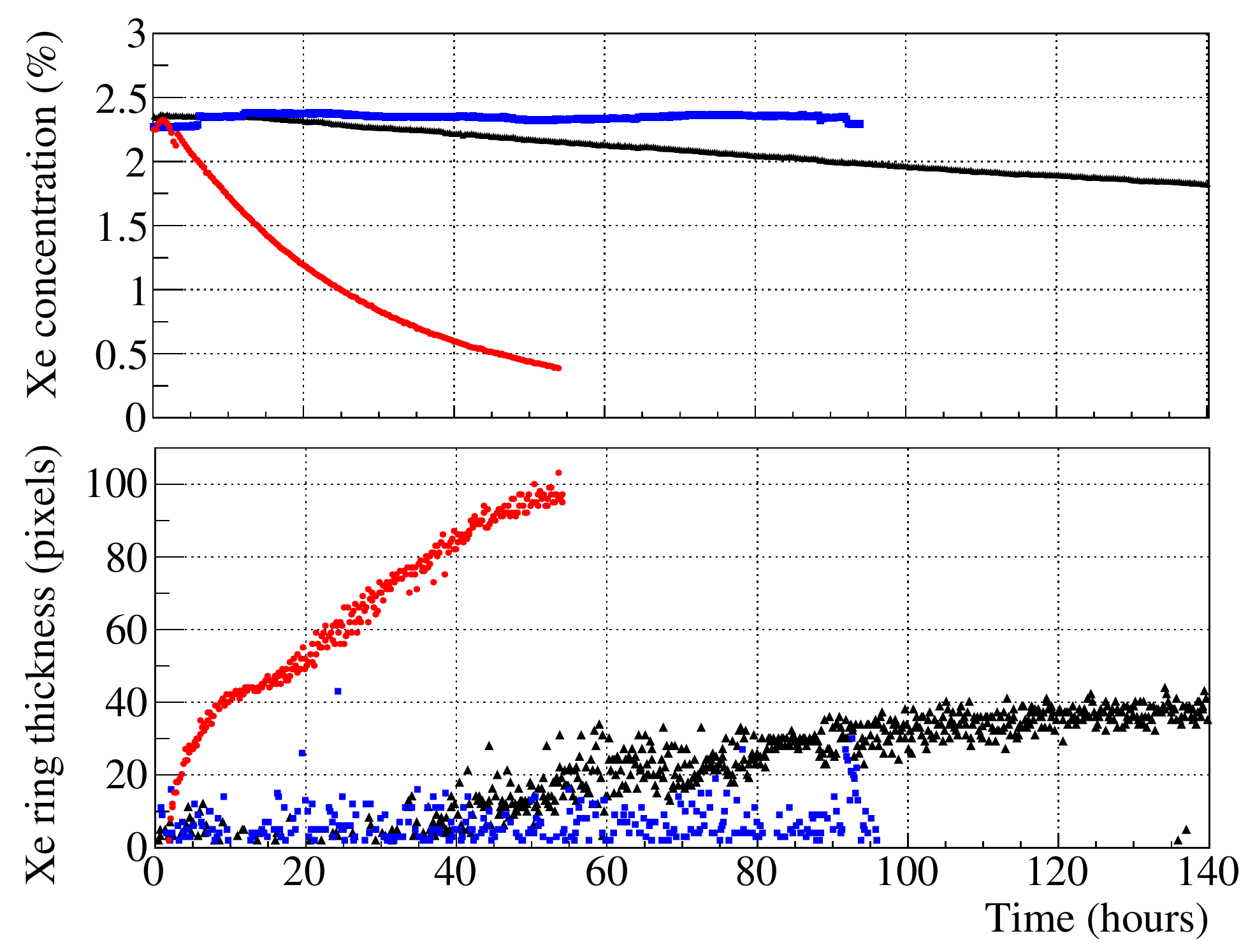}
\caption{Time dependence of the xenon concentration in the liquid mixture measured with the capacitor meter (top) and xenon ice accumulation on the detector walls (bottom) during the 3 stability tests: blue -- near-zero thermal gradient between the top and the bottom of the detector, red -- large thermal gradient (75~K), black -- small thermal gradient (10K). The ice ring image thickness values in the zero-gradient test (blue squares in bottom figure) are nearly all analysis artifacts due to the absence of ice rings. }
\label{fig:stability}
\end{figure}

\begin{figure}[!t]
\centering
\includegraphics[width=0.49\textwidth]{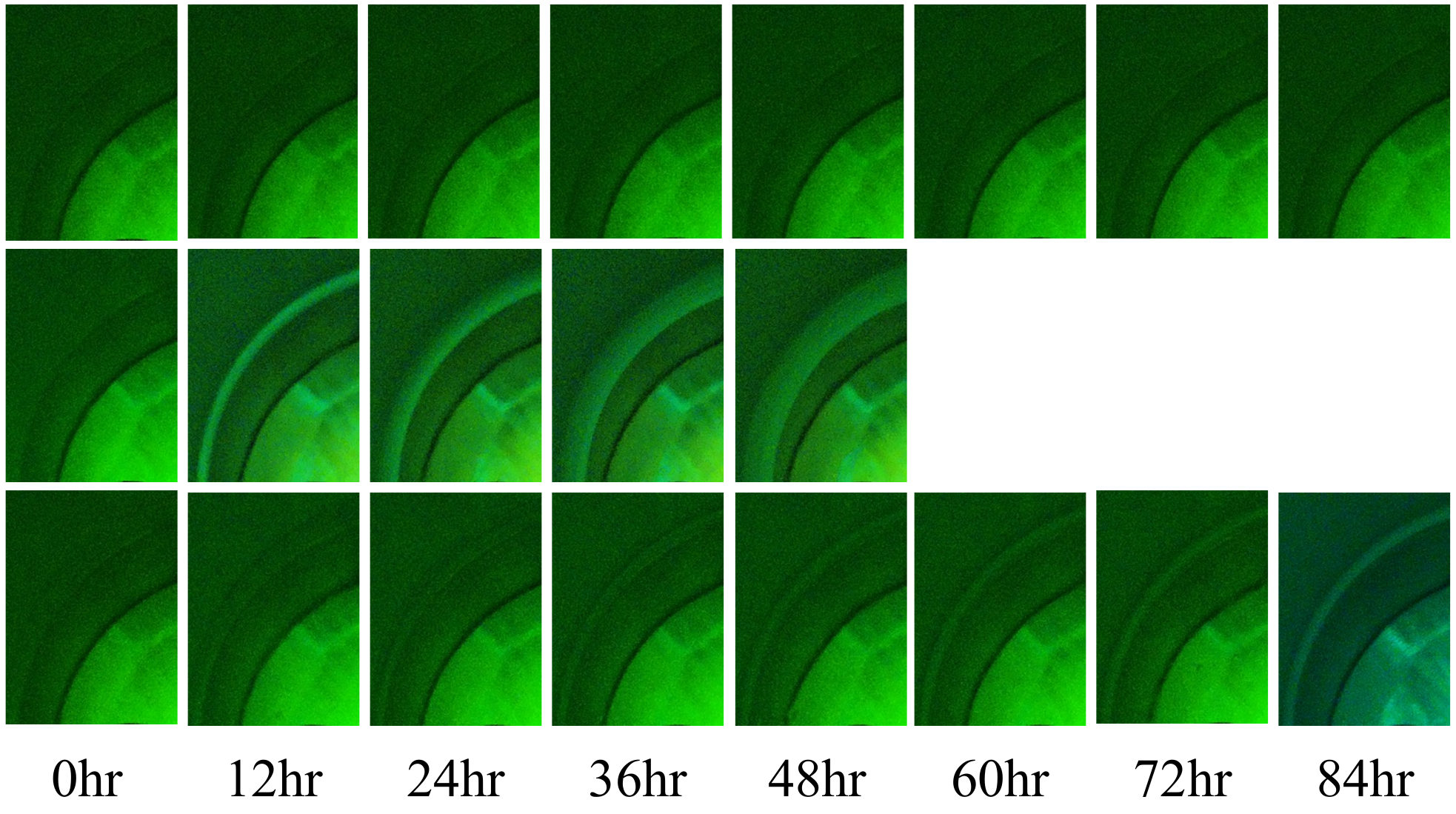}
\caption{Images of the detector bath near the liquid surface at different times from the beginnings of test 1 (top row), test 2 (middle row), and test 3 (bottom row). The bright ring features in the middle and bottom rows are xenon ice and the liquid surface is a few mm below the ring position. }
\label{fig:imageevolution}
\end{figure}

Figure~\ref{fig:stability} (top, blue curve) shows the measured xenon concentration inside the detector volume over a period of 4 days in this first test.  The concentration was stable at 2.35$\pm$0.05\% throughout this period, and the observed variations are attributed to electrical noise in the capacitance measurement system.  Images of the liquid surface and the nearby detector walls at different times during the test are shown in Fig.~\ref{fig:imageevolution}, and no changes are observed. 

Additional tests were carried out in which significant thermal gradients between the detector flange and the liquid volume were allowed. 
In the second test, we removed gas from the TSU and did not actively control the temperature of the detector flange. In this scenario, the detector flange receives heat from detector components connecting it to the room-temperature vacuum flange and from radiation, while it is also cooled by upward flowing cold argon gas and by thermal conduction through the detector can toward the liquid.  The detector flange equilibrated at 168--170~K, producing a significant heat flow 
through the detector walls toward the detector bath.  
Over a testing period of 2 days, the xenon concentration decreased continuously to approximately 0.4\%. 
Meanwhile, we observed a slow, continuous growth of xenon ice on the inner surface of the detector wall and the coaxial cables serving the capacitive concentration meter. The ice began as a thin ring a few mm above the liquid and grew to several mm in height and thickness while significantly overhanging the liquid surface, as illustrated in Fig.~\ref{fig:imageevolution} (second row). 
The xenon ice ring appeared to have the same size and shape on opposite sides of the detector can and the rings around different cables were similarly matched in appearance.  
We estimated the thickness of the ice ring {\it image} (at around 11 o'clock in the camera view) by measuring its width in image pixels as a function of time.  The result is shown in Fig.~\ref{fig:stability} (bottom). 
Ice was notably absent on the thin G10 laminate board strip of the liquid level meter, which is a poor thermal conductor.
All of these observations confirm heat flow into the liquid as the driver of the mixture instability.

In the third test, the temperature of the detector flange was maintained at 10~K above detector bath by the TSU, mimicking a detector that regulates the temperature profile but fails to do so precisely. 
Similar to the second test, the xenon concentration decreased over time but with a reduced slope, and in this condition a xenon ice ring also grew above the liquid surface but at a slower rate. 
Results from the second and third tests empirically suggest a xenon ice formation rate that increases with the concentration of dissolved xenon and the power delivered conductively to the liquid surface, consistent with the simple distillation mechanism explained in Section~\ref{sec:instability}.

\begin{figure}[!t]
\centering
\includegraphics[width=0.4\textwidth]{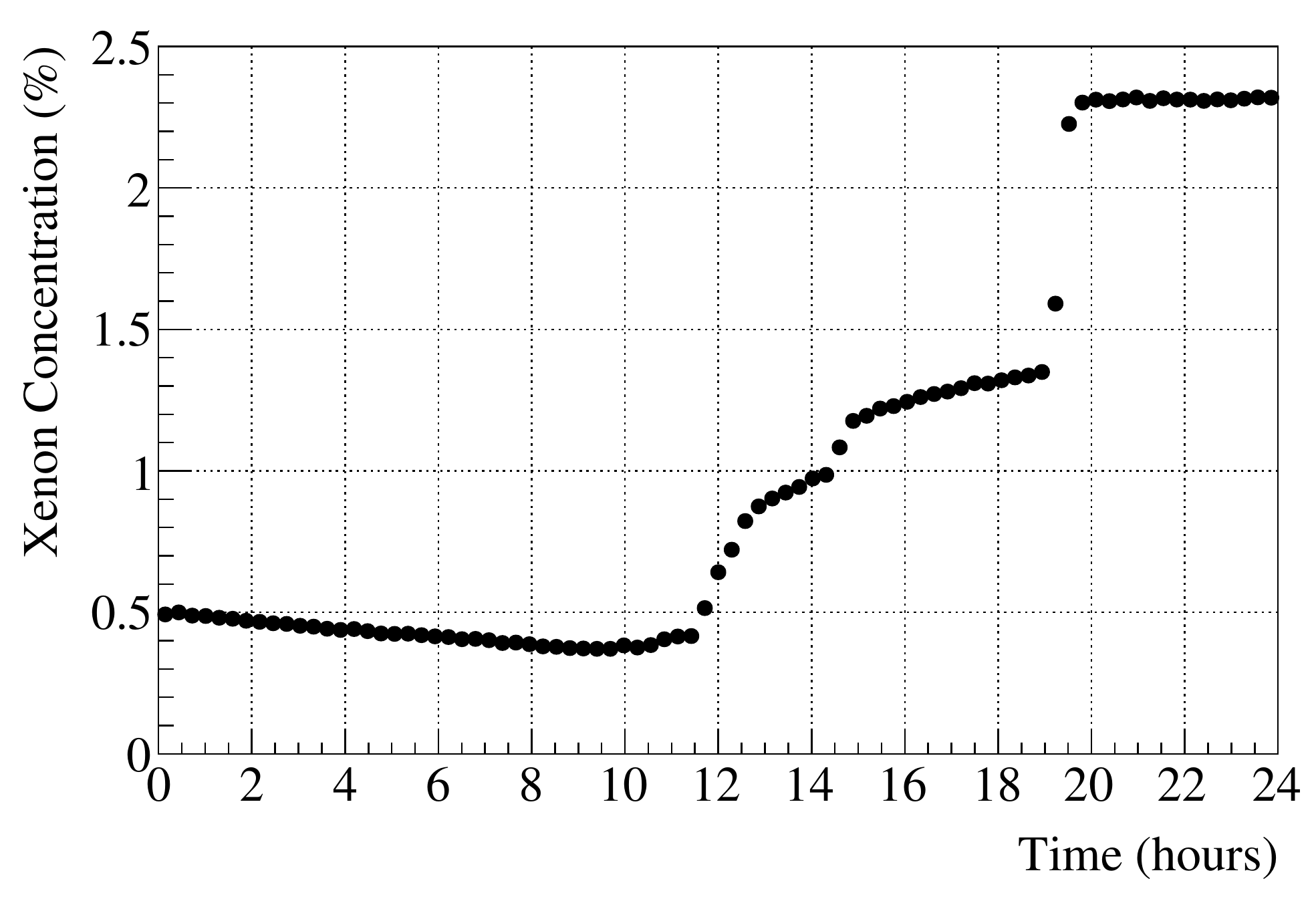}
\caption{Recovery of xenon concentration in the liquid mixture following the second test by condensing argon gas on the top detector flange, beginning near hour ten. }
\label{fig:xerecovery}
\end{figure}

We also tested the condition where the detector flange temperature was set below the temperature of the detector bath. As described in Sec.~\ref{subsec:doping}, argon gas can condense on the detector flange, and the downward flowing liquid argon film dissolves xenon ice from the detector walls and move xenon back into the detector bath.
Figure~\ref{fig:xerecovery} shows the measured xenon concentration in the liquid mixture after the second test, when the detector top flange was cooled to about 1 degree below the liquid bath temperature. 
Gradual increases of the xenon concentration were a result of xenon ice dissolving in the downward argon film flow, and the abrupt increases coincided with large pieces of xenon ice falling into the liquid. 
Within 10 hours the xenon concentration in the liquid mixture was restored to the level measured prior to the second test. 
This result suggests that xenon ice buildup on the detector wall is responsible for nearly all of the xenon segregating from the liquid mixture. This procedure was performed after each stability test discussed above to reset the initial condition. 
Note that operating the system in this sub-cooling mode slows or stops the liquid flow from the HX to the detector, and extended operation in this mode can lead to an increase of xenon concentration inside the HX and depletion from the detector bath. 

\subsection{Implications for future xenon doping efforts}
\label{subsec:discussion}

The concept of doping liquid argon with a small fraction of xenon to improve its performance as a scintillation and/or ionization detection medium has attracted broad interests. 
This work provides an extensive experimental study of the instability modes that may develop in a xenon-doped argon system during operation in steady-state circulation and xenon introduction stages. 
Although the experiments were performed at high doping concentrations, the instabilities explored have implications for all future doping efforts.

In ideal situations, xenon-doped liquid argon is stable until the xenon concentration reaches the solubility limit. 
In reality, inhomogeneity of xenon concentration in a xenon-doped liquid argon detector can develop at much lower xenon doping levels. 
The primary cause of this instability is phase changes of the mixture. 
A net evaporation of the liquid mixture creates a distillation scenario that increases the xenon concentration in the liquid; in extreme cases, the local xenon concentration can rise to above the solubility limit and ice will form, which can lead to a mechanical or electrical failure of the detector.   
Similarly, condensation removes xenon from both the gas and the liquid by creating a solid phase if the introduced xenon partial pressure is above the saturation vapor pressure at the condensation temperature. 

Therefore a xenon-doped liquid argon detector should avoid unnecessary and unintended phase changes to improve the system stability. 
For example, if a detector can directly purify the liquid~\cite{Aprile2022_LiquidPurification} instead of evaporating the liquid into a gas for purification, the main distillation instability may be avoided entirely. 
As demonstrated in this work, argon can still be circulated in the gas phase by evaporating directly from the detector bath when the system condition is properly controlled. 
In this case, an induced liquid convection beneath the evaporating surface prevented over-concentration of xenon at the evaporation surface and stabilized the detector. 
However, this simple method only extracts and purifies argon and other gas components of similar or higher vapor pressure while much less volatile species such as water are unable to leave the liquid.  An alternative purification scheme is to draw the liquid mixture into an isolated hot chamber where the temperature is maintained above the xenon triple point. There the liquid boils completely and no xenon residual accumulates so no distillation occurs.  This implementation requires careful engineering so that evaporation only takes place inside the hot chamber; earlier evaporation in cold liquid transfer lines will result in the familiar problems of xenon concentration, ice formation and possibly clogging.

In lightly doped systems with large detector liquid volumes, such as a large liquid argon TPC, the traditional circulation scheme may be used in which the liquid mixture is evaporated in a small evaporation volume and the produced gas is fed into the circulation path. 
In this design, the xenon concentration within the liquid at the point of evaporation simply rises through distillation until the xenon concentration of the evaporated gas matches the doping level of the liquid in the detector.  In this equilibrium state, the xenon concentration in the detector liquid is a factor of $H$ (Henry's constant, see Sec.~\ref{subsec:solubility}) lower than that of the liquid at the point of evaporation. For a system operating at one atmosphere, the saturation limit of 5\% and $H\simeq~1240$ set a theoretical maximum of 40~ppm for the liquid doping level, which is sufficient to provide useful energy transfer \cite{Cheshnovsky1972} \cite{Neumeier2015A}.
However, special precaution should be taken for the design of the evaporation volume where the xenon concentration is near the solubility limit and instability can develop. 

Unintended phase changes can occur in a detector system and their mitigation is more challenging. 
Variations in either temperature or pressure can locally displace the liquid surface from the vapor curve and provoke phase changes. 
The first aspect of preserving the system stability is local control of heat flux, with the expectation that heat flux into the liquid results in enhanced evaporation and heat flux out of the gas results in condensation.  The second aspect is to mechanically design the system so that where heat flux occurs natural convection can relieve the local temperature and xenon concentration excursions. The flow of xenon within the liquid is strongly analogous to that of heat.  The diffusion constant of xenon in liquid argon is $\mathcal{O}(100)$ times lower than the thermal diffusivity of liquid argon, and at the length scale of our detector both diffusive flows are strongly subdominant to those of convection~\cite{ArLiqSelfDiff, NISTwebbook}.  Note that convection is primarily driven by density changes from thermal expansion, not increased xenon concentration.  Consequently, designs that yield good thermal homogeneity will also tend to yield good xenon homogeneity.
As discussed in Sec.~\ref{sec:instability} and demonstrated in Sec.~\ref{subsec:stabilitytest},
unwanted ice formation is most likely to occur where heat is introduced to portions of the liquid that cannot mix convectively with the bulk of the bath, which can be mitigated by controlling heat and liquid flows.  A similar phenomenon of ice formation was observed above a charge amplifier mounted to a horizontally oriented circuit board submerged just below the liquid surface \cite{XinranLi2020}.  The board geometry impeded convective mixing of the concentrated liquid resulting from the heat input of the charge amplifier.  

Certain phase changes cannot be avoided in a detector. 
In a xenon-doped argon scintillating bubble chamber, the liquid mixture has to undergo frequent overheating and cooling phases in order for the system to operate, so the resulting inhomogeneity in the xenon concentration and its mitigation have to be carefully considered. 
Beside, at the xenon doping stage of an experiment, the introduced xenon gas will have to be condensed and mixed with liquid argon. 
In this work, we have shown that direct condensation of argon-xenon gas mixtures with xenon partial pressures well above the xenon vapor curve is feasible.  This was achieved by introducing the gas mixture into an actively cooled liquid argon bath while preventing xenon from freezing before it enters and mixes in the liquid volume.  This approach can significantly reduce the time or flow rate required for xenon introduction in a detector that requires large quantities of xenon.

This work also demonstrates that xenon frost or ice attached to surfaces above the liquid surface can be returned to the detector bath without emptying and warming the detector.  As discussed in Sec.~\ref{subsec:stabilitytest}, when detector components above the liquid are cooled to below the liquid temperature, argon condensed on these cold components dissolved solid xenon while flowing toward the bath.  Although this procedure recovered nearly all of the xenon in this work, this may not be the case for more complex detector geometries that cannot be flushed by the liquid argon reflux.  In those cases, alternative means of wetting the detector structures might be used, such as a rapid step in pressure (to force condensation), temporarily raising the liquid level, agitating the liquid surface, or  spraying the exposed structures.

Finally, diagnostic tools are necessary to continuously monitor the xenon concentrations at critical locations of a system. The capacitance meter used in this work achieved a sensitivity of 1~fF with a commercial readout meter, which corresponds to 0.05\% of xenon doping in liquid argon. The sensitivity can be improved by using larger electrodes and upgraded capacitance measurement methods. 
However, to measure ppm-level concentrations, a more practical approach may be to design the cryogenic system to allow a liquid sample to be isolated and completely evaporated, and then to measure the partial pressures argon and xenon in the gas.  The SRS RGA200 system used in this work is capable of measuring ppm-levels of xenon in argon gas, despite difficulties in performing robust measurements.

\section{Conclusion}
\label{sec:concl}

We designed and constructed a system to investigate thermodynamic instabilities associated with xenon doping in liquid argon. 
The apparatus is capable of condensing xenon-rich argon gas at the percent level and stabilizing the resulting liquid argon-xenon mixture with over 2\% xenon doping ($\sim 3$ times below saturation limit) for several days with no evidence of degradation.
For contrast, we also demonstrated that instability in these mixtures develops when the thermal profile of the system is not accurately controlled, which caused xenon to concentrate, solidify, and separate from the liquid mixture. 
This work develops the knowledge and tools to investigate instability modes of proposed xenon doping in large liquid argon detectors.
Additionally, it demonstrates the feasibility of operating a dual-phase argon detector with heavy xenon doping to enhance its sensitivity to low energy ionization signals such as those expected from low mass dark matter interactions and the nuclear scattering of low energy antineutrinos from nuclear reactors.

\section{Appendix}
\label{sec:appendix}

\subsection{The solubility and partial pressure of xenon argon mixtures}
\label{subsec:solubility}

Following the review by Tegeler {\it et al.} \cite{tegelerArEqofState} of precision measurements of pure argon, we adopt the vapor pressure fit determined by Gilgen {\it et al.} \cite{gilgenArVapP}:

\begin{equation}
\ln{\bigg{(}\frac{P_{sAr}}{P_{cAr}}\bigg{)}} = \frac{T_{cAr}}{T}(a_1 q + a_2 q^{3/2} + a_3 q^{2}+ a_4 q^{9/2})
\end{equation}

where $P_{sAr}$ is the saturated vapor pressure, $P_{cAr} = 4.863 \mathrm{MPa}$ is the critical pressure, $T_{cAr} = 150.687 {\mathrm K}$ is the critical temperature, ${\mathrm T}$ is the temperature, $q = 1-T/T_{cAr}$, $a_1 = -5.9409785$, $a_2 = 1.3553888$, $a_3 = -0.46497607$ and $a_4 = -1.5399043$.\\

The vapor pressure of solid xenon is approximated by a fit to measurements over the range 76.2~K to 104.0~K as
\begin{equation}
\label{Xesolid_vap_P}
\ln{P_{sXe}} = a / T + b    
\end{equation}
where $a$ = -1960.37~K and $b$ = 18.9607 \cite{Xe_Vap_P_sol_Leming_Pollack}.

The saturation limits and vapor pressures of xenon-doped argon mixtures were measured near atmospheric pressure by Yunker and Halsey \cite{Yunker1960}.  The saturation limit is given by
\begin{equation}
\label{eqn:satlimiteq}
\ln{(n_{Xe}^{sat})} = 1.6463 - 406 {\mathrm K} / T
\end{equation}

\begin{figure}[!t]
\centering
\includegraphics[width=0.45\textwidth]{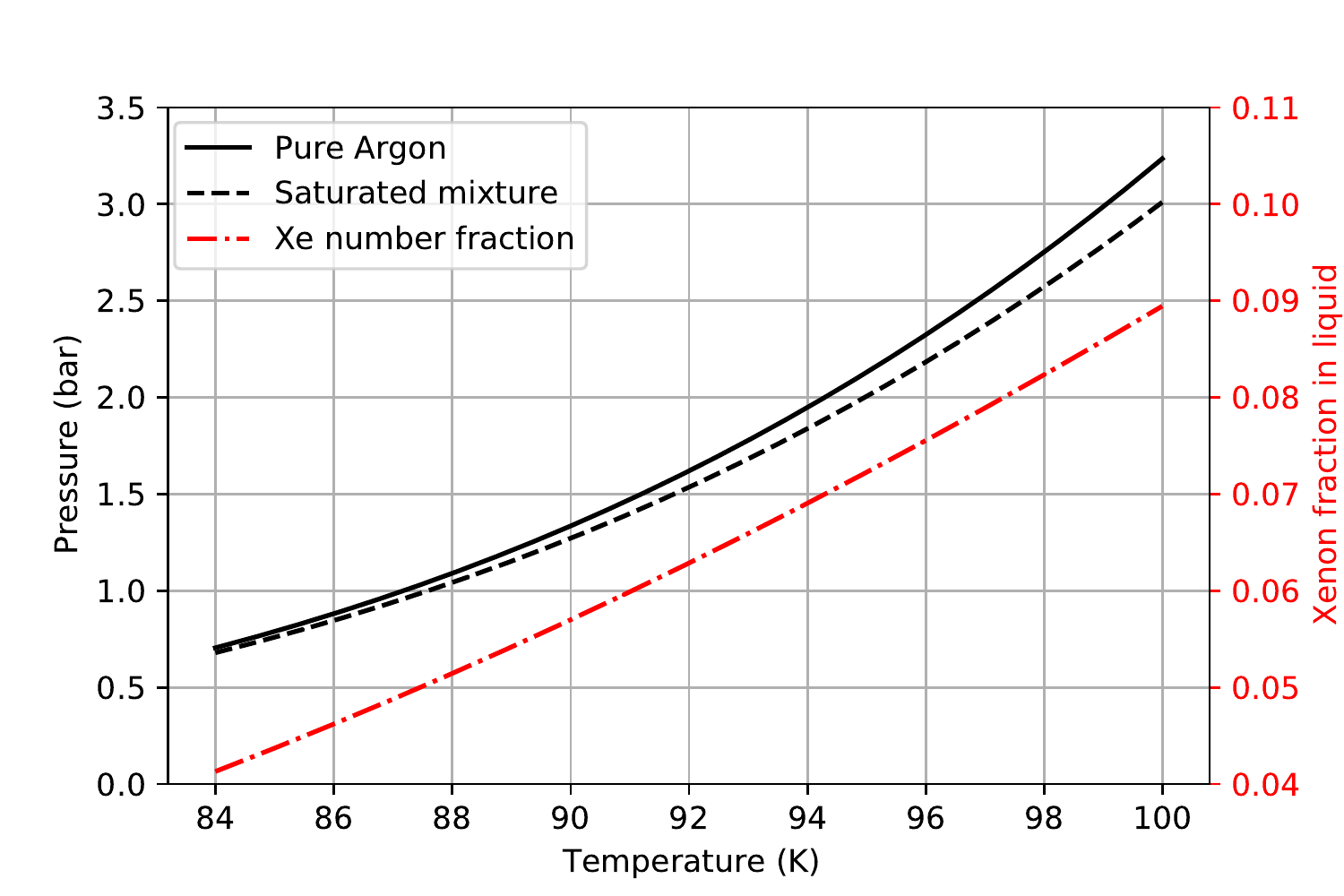}
\includegraphics[width=0.45\textwidth]{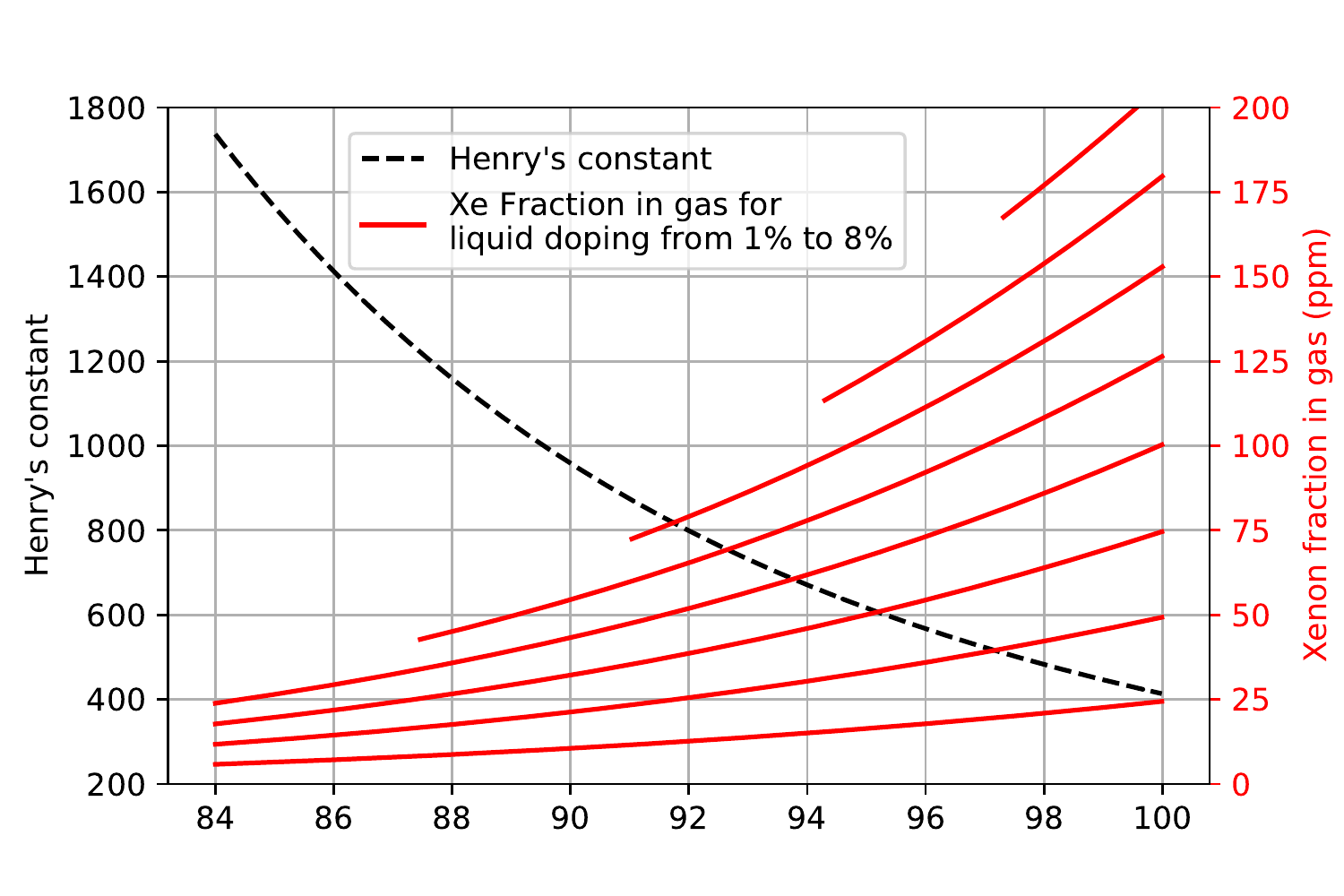}
\caption{{\bf Top:} The vapor pressure of pure liquid argon (solid line) and liquid argon saturated with xenon (dashed line) as a function of temperature; the solubility limit of xenon in liquid argon is also shown (dot-dash line, scale on the right axis); {\bf Bottom:} Henry's constant at different temperatures (dashed line), and the xenon concentration in the gas phase above a liquid with 1--8\% (from bottom to top) xenon doping by mol fraction (scale on the right axis).}
\label{fig:sathenry}
\end{figure}

The vapor pressure above such mixtures is almost entirely (within $\mathcal{O}(10^{-4})$) due to argon, 
which is depressed relative to that of pure argon by Raoult's law.  An additional correction to the vapor pressure is due to the larger attractive forces among xenon atoms in the liquid mixture, giving
\begin{equation}
\label{Pdepression}
\ln{\bigg{(}\frac{P_{mixAr}}{P_{sAr}}\bigg{)}} = \ln{(1-n_{Xe})} + \alpha n^2_{Xe}/T    
\end{equation}
where $n_{Xe}$ is the xenon mole fraction and $\alpha = 269.2 {\mathrm K}$ \cite{Yunker1960}.  The effect of Raoult's law is about ten times the magnitude of the $\alpha$ term at 4\% xenon concentration.  The same work also demonstrates that the solid phase that forms at the saturation limit is at least 99.5\% xenon and that the mixture is well modeled as a regular solution.  The vapor pressure predicted by equation \ref{Pdepression} closely matches a report of 8.45 bar measured above a 12.5\% mixture at 116~K \cite{xenonmixtwelvepercent}. 
Our evaluated vapor pressure values for pure argon and saturated argon-xenon liquid mixtures as a function of temperature are shown in Fig.~\ref{fig:sathenry} (top); the xenon solubility limit from equation \ref{eqn:satlimiteq} is also drawn. 

The evaluation of xenon partial pressure above an argon-xenon mixture is more complex. 
At the point of xenon saturation onset three phases exist, leaving one degree of freedom (either pressure or temperature), as specified by the Gibbs phase rule.  At saturation onset and thermodynamic equilibrium the xenon chemical potential $\mu_{Xe}$ is equal among the three phases.  Assuming the solid phase is pure xenon and neglecting any interactions between xenon and argon in the gas phase, the xenon vapor pressure above the mixture at the saturation limit is given by $P_{sXe}$~\cite{mixturechembook1952}.  This holds even if no solid phase is exposed to the gas phase.  
As for unsaturated mixtures, no experimental measurement or numerical simulation of the xenon partial pressure above the mixture exists to our best knowledge, so we can only estimate the xenon partial pressure by scaling with Raoult's law.  That is,

\begin{equation}
\ln{\bigg{(}\frac{P_{mixXe}}{P_{sXe}}\bigg{)}} \simeq  \ln{\bigg{(}\frac{n_{Xe}}{n_{Xe}^{sat}}\bigg{)}}    
\end{equation}
 
This likely underestimates the actual xenon partial pressure for unsaturated mixtures because their attraction to a xenon atom is less than that of a saturated mixture at the same temperature.  
Still, this calculation predicts a much larger (by a factor $1/n_{Xe}^{sat}$) partial pressure of xenon gas above the saturated liquid than is estimated by multiplying the vapor pressure of solid xenon at the mixture temperature by $n_{Xe}$.  
As suggested in the original solubility experiment \cite{Yunker1960}, argon-xenon solutions are regular but not ideal, based on the fact that xenon is not completely miscible in argon at the temperatures of interest. 
For dilute regular solutions, the vapor pressure of the solute (xenon) follows Raoult's law but is shifted by a term accounting for the different energies of a solute molecule within an environment of pure solute versus a solute molecule within an environment of (nearly) pure solvent \cite{mixturechembook1952}.  Since the vapor pressure over pure xenon matches that over the mixture at the point of saturation, the upward shift in vapor pressure is $1/n_{Xe}^{sat}$.    

In this case, an estimate of the dimensionless Henry's law constant is
\begin{equation}
H = \lim _{n_{Xe} \to 0} \frac{(P_{mixAr}+ P_{mixXe})\cdot{n_{Xe}}}{P_{mixXe}}
\end{equation}
which measures the suppression of xenon mole fraction in the vapor relative to that in the liquid mixture. 
Our calculated Henry's constant and the estimated xenon mole fraction in the gas at different temperatures and different xenon concentrations in the liquid are shown in Fig.~\ref{fig:sathenry} (bottom).

\subsection{Dielectric constant dependence on xenon-doping}
\label{subsec:dielectricDependence}

As discussed in \ref{subsec:capacitancemeter}, a parallel plate capacitor is implemented to measure the xenon concentration in the liquid mixture. To convert from capacitance to xenon concentration, a calibration function must be fit from data. One can determine an appropriate function model to fit for the dependence of the argon-xenon fluid's dielectric constant $\epsilon_r$ on the xenon molar fraction from the Clausius-Mossotti relation:

\begin{equation}
\frac{\epsilon_{r}-1}{\epsilon_{r}+2} = \sum_{i=1}^2 \frac{n_i\alpha_i}{3\epsilon_{0}}
\end{equation}

Where $n_i$ is the number density of molecule (or atom) type i, $\alpha_i$ is the polarizability of a molecule type i, and $\epsilon_0$ is the permittivity of free space. In this application, $i=1$ corresponds to argon, and $i=2$ corresponds to xenon. Henceforth, we will write $n_1$ as $n_{Ar}$ and $n_2$ as $n_{Xe}$.  One can express $n_i$ as the ratio of the total number of atoms of type i (which we will refer to as $N_i$) and volume of the bath V. Let us also assume, given a constant $N_{Ar}$, that V is a linear function of $N_{Xe}$ in the regime of doping concentrations below the saturation limit. Finally, one can substitute $N_{Xe}$ with a more useful quantity: $F_{Xe}$, which denotes the molar fraction of xenon in the bath. For a constant $N_{Ar}$ the dielectric constant can be isolated and expressed as a single-variable function of $F_{Xe}$:

\begin{equation}
\label{eqn:epsDep}
\epsilon_r(F_{Xe}) = \frac{A+BF_{Xe}}{C+DF_{Xe}}
\end{equation}

Where $A$,$B$,$C$, and $D$ are constants with respect to $F_{Xe}$:

\begin{align*}
\label{eqn:parABCD}
& A = 3\epsilon_{0}b+2N_{Ar}\alpha_{Ar}\\
& B = 3\epsilon_{0}N_{Ar}m+2\alpha_{Xe}N_{Ar} - 2\alpha_{Ar}N_{Ar} - 3\epsilon_{0}b\\
& C = 3\epsilon_{0}b-N_{Ar}\alpha_{Ar}\\
& D = 3\epsilon_{0}N_{Ar}m-\alpha_{Xe}N_{Ar} + \alpha_{Ar}N_{Ar} - 3\epsilon_{0}b\\
\end{align*}

Where $m$ and $b$ are parameters of the linear function $V(N_{Xe})$. For small $F_{Xe}$ one can apply the binomial approximation to the denominator of \ref{eqn:epsDep}, yielding a quadratic dependence of $\epsilon_r$ on $F_{Xe}$. For small $F_{Xe}$ the quadratic term is dominated by the linear term. Dropping the quadratic term reveals an approximately linear dependence of $\epsilon_r$ on $F_{Xe}$. This is our justification for modeling the change in dielectric constant (and therefore the change in the difference between the active and reference capacitors' capacitance) as a linear function of $F_{Xe}$.

\section*{Acknowledgments}

This work was performed under the auspices of the U.S. Department of Energy by Lawrence Livermore National Laboratory (LLNL) under Contract DE-AC52-07NA27344 and was supported by the LLNL-LDRD Program under Project No. 20-SI-003 (LLNL release number LLNL-JRNL-839368).
J. Kingston and R. Smith are partially supported by the DOE/NNSA under Award Number DE-NA0000979 and DE-NA0003996 through the Nuclear Science and Security Consortium. 
C. G. Prior is partially supported by the DOE under Award Number DE-SC0010007. 

We thank Xinran Li from Princeton (now at LBNL) for discussions about capacitance measurements of liquid argon and the thermodynamic stability of xenon argon mixtures.
We thank Nathan Eric Robertson, Phillip Hamilton, and Sean Durham from LLNL for their technical support on fabricating parts for the liquid argon detector. 

%-----------------------------------------------------------------------
%-----------------------------------------------------------------------
\bibliography{main}
%\nocite{*}
\bibliographystyle{apsrev}
\end{document}